\documentclass[prl,twocolumn,showpacs,floatfix,superscriptaddress,amsmath,amssymb]{revtex4}
\usepackage{dcolumn}% Align table columns on decimal point
\usepackage{bm}% bold math
\usepackage{natbib}

\usepackage{pstricks}
\usepackage[latin1]{inputenc}
\usepackage{amsmath}
\usepackage{amssymb}
\usepackage{graphicx}
\usepackage{graphics}
\usepackage{color}
\def\vr {{\bf r}}

\def \ba {\begin{eqnarray}}
\def \ea {\end{eqnarray}}
\newcommand{\nn}{\nonumber}

\newcommand{\op}[1]{\mathbf{#1}}

\def\prl#1#2#3{Phys.\ Rev.\ Lett.\ {\bf #1}, #2 (#3)}

\def\prb#1#2#3{Phys.\ Rev.\ B {\bf #1}, #2 (#3)}

\newcommand{\sumauno}{%
\raisebox{0.0ex}{\includegraphics[scale=0.15]{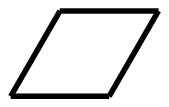}}
}

\begin{document}

%\title{Phase transitions induced by magnetic field in the classical $J_1-J_2$ antiferromagnetic Heisenberg model
%on honeycomb lattice.}
% \title{Emerging discrete broken symmetries in a 2D frustrated Heisenberg model}
\title{Broken discrete symmetries in a frustrated honeycomb antiferromagnet}

\author{H. D. Rosales}
\affiliation{IFLP, Departamento de F\'isica, Universidad Nacional de La Plata, La Plata, Argentina}
% \affiliation{Commissariat \`{a} l'Energie Atomique, DSM/INAC/SPSMS,
% F-38054 Grenoble, France}

\author{D. C. Cabra}
\affiliation{IFLP, Departamento de F\'isica, Universidad Nacional de La Plata, La Plata, Argentina}

\author{C. A. Lamas}
\affiliation{Laboratoire de Physique Th\'eorique,  IRSAMC,
 CNRS and Universit\'e de Toulouse, UPS, F-31062 Toulouse, France}

\author{P. Pujol}
\affiliation{Laboratoire de Physique Th\'eorique,  IRSAMC,
 CNRS and Universit\'e de Toulouse, UPS, F-31062 Toulouse, France}

\author{M. E. Zhitomirsky}
\affiliation{Service de Physique Statistique, Magn\'etisme et Supraconductivit\'e,
UMR-E9001 CEA-INAC/UJF, 17 rue des Martyrs, 38054 Grenoble Cedex 9, France}

\date{\today}

%%%%%%%%%%%%%%%%%%%%%%%%%%%%%%%%%%%%%%%%%%%%%%%%%%%%%%%%%%%%%%%%%%%%%%%%%%%%%%%%
\begin{abstract}
We study the magnetic phase diagram of the $J_1$--$J_2$ Heisenberg antiferromagnet
on a honeycomb lattice at the strongly frustrated point $J_2/J_1=1/2$ using large-scale
Monte Carlo simulations. At low temperatures we find three different field regimes,
each characterized by different broken discrete symmetries.  In low magnetic fields
up to $h_{c1}/J_1\approx 2.9$ the $Z_3$ rotational lattice symmetry is spontaneously
broken while a 1/2-magnetization plateau is stabilized around $h_{c2}/J_1=4$.
The collinear plateau state and the coplanar state in higher fields break
the $Z_4$ translational symmetry and correspond to triple-$q$ magnetic structures.
The intermediate phase $h_{c1}<h<h_{c2}$  has an interesting symmetry structure,
breaking simultaneously the  $Z_3$ and $Z_4$ symmetries.
At much lower temperatures the spatial broken discrete symmetries coexist with the quasi
long-range order of the transverse spin components.
\end{abstract}
%%%%%%%%%%%%%%%%%%%%%%%%%%%%%%%%%%%%%%%%%%%%%%%%%%%%%%%%%%%%%%%%%%%%%%%%%%%%%%%%

%\pacs{}
\maketitle

The search for a quantum spin liquid  -- an insulating magnet with a gapless ground state
which breaks neither lattice nor spin symmetries -- has been the focus of many studies on two-dimensional
frustrated quantum antiferromagnets \cite{Wang,Meng,Cabra,Farnel}.
Such systems are assumed to be the main candidates to describe  a
rich variety of unconventional phases, phase transitions and critical points with
deconfined fractional excitations \cite{Anderson,Fazekas,Raman}. Frustration 
plays an important role in classical systems as well. Within this context, the phenomenon of order
by disorder (OBD) \cite{OrderbyDis} is the perfect example where the interplay
of frustration and fluctuations produces the emergence of unexpected order. OBD implies that
certain low-temperature spin configurations are favored by  higher entropy rather than by lower energy.
%%%%%%%%%%%%%%%%%%%%%%%%%%%%%%%%%%%%%%%%%%%%%%%%%%%%%%%%%%%%%%%%%%%%%%%%%%%%%%%%%%%%%%%
For instance, some frustrated spin models may exhibit magnetization plateaus even at
the classical level \cite{Zhitomir,LhuiMis}. In this case fluctuations are responsible
for stabilizing particular collinear spin configurations that have softer excitation
spectra compared to a general noncollinear spin state. Another nontrivial fluctuation
effect is a finite-temperature transition in 2D Heisenberg antiferromagnets related
to breaking (discrete) lattice symmetries in the absence of a long-range magnetic
order \cite{Chandra90}. Studying various fluctuation-induced types of magnetic order is
important in order to establish robustness of a hypothetical spin-liquid state.

\begin{figure}[t!]
\begin{center}
\includegraphics[height=8.5cm]{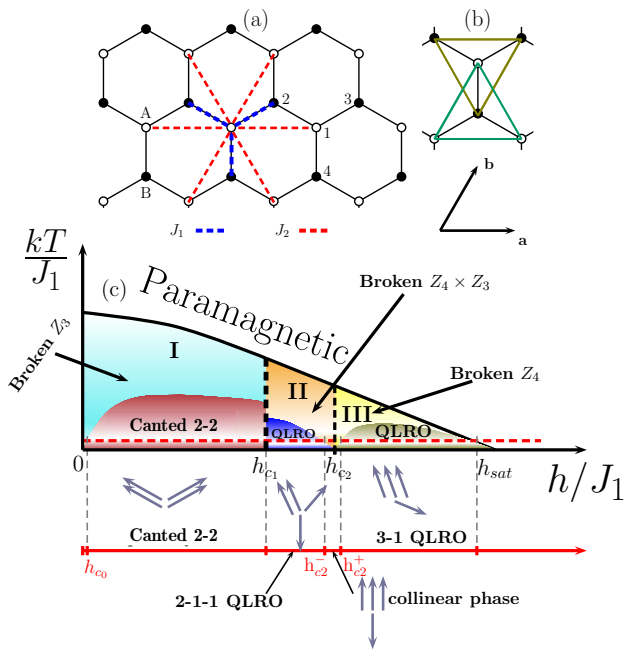}
\end{center}
\caption{(color online) (a) Honeycomb lattice. Blue (Red) line represents first (second)  nearest
neighbors. (b) Elementary tetrahedrons used to write the Hamiltonian  in the highly frustrated
point. (c) Schematic phase diagram. The dotted line represent a constant low temperature line
which can be obtained by increasing the magnetic field. By moving along this line one should observe phases I, II and III with broken $Z_3$, $Z_3 \times Z_4$ and $Z_4$ symmetries
respectively. In phase (III) the lattice orientational symmetry is restored. If the temperature is low enough (as depicted in the figure) QLRO phases should be present everywhere except in the vicinity of $h=0$ and $h_{c_2}/J_1 =4$ where the pseudo plateau with a collinear configuration is present. Finally, increasing further the magnetic field should drive the system into the trivial paramagnetic phase characteristic of the high temperature region.}

%Phase I  corresponds to a broken $Z_{3}$ symmetry. In this
%phase translational and rotational symmetries of the lattice are simultaneously broken.
%Raising the value  of the magnetic field above $h_{c_{1}}$ at fixed temperature, a 2-1-1 phase is
%observed (phase II). This phase has broken $Z_3 \times Z_4$ symmetry. By increasing further the magnetic field this phase transforms continuously into the uuud structure around  $h_{c_2}/J_1 \approx 3.7$
%in which the orientational symmetry is restored when the system enters in  phase III, which
%corresponds to a broken $Z_{4}$  symmetry. This phase has a collinear order in the direction
%of the magnetic field and it is disordered in the $XY$ plane.}
\label{fig:PhaseDiagram}
\end{figure}

Here, we consider the Heisenberg  antiferromagnet on a honeycomb lattice, model
realized in a number of real magnetic materials \cite{exp_1,exp_2,exp_4,Ressouche10}.
Recent interest in this model is largely motivated by the experimental realization
of the spin-liquid state in $\rm Bi_3Mn_4O_{12}(NO_3)$ \cite{Matsuda10}.
Mulder {\it et al.}\ \cite{Mulder2010} have studied the $J_1$--$J_2$ frustrated 
honeycomb antiferromagnet,  Fig.~\ref{fig:PhaseDiagram}(a), in zero magnetic field.
They have found that quantum fluctuations select a family of special states
characterized by three inequivalent ${\bf Q}$ vectors for $1/6<J_2/J_1\leq 1/2$ and 
by three different ones for $1/2<J_2/J_1$. While at finite temperatures thermal 
fluctuations melt the spiral order, the discrete $Z_3$ lattice rotational symmetry 
is still broken at low temperatures.

In this Letter, we extend the previous theoretical work on
the $J_1$--$J_2$  honeycomb antiferromagnet to finite magnetic fields.
This is important in view of the experimentally observed field-induced transition
between the spin-liquid state and a long-range ordered magnetic structure
in  $\rm Bi_3Mn_4O_{12}(NO_3)$ \cite{Matsuda10}.
Specifically, we focus on the highly frustrated point at $J_2/J_1=1/2$.
Surprisingly for such a simple model, we find a plethora of new emergent
broken symmetries, which may exist in magnetic fields alongside with a
1/2-magnetization plateau.
%%%%%%%%%%%%%%%%%%%%%%%%%%%%%%%%%%%%%%%%%%%%%%%%%%%%%%%%%%%%%%%%%%%%%%%%%%%%%%%%%%%%%%%%%%%%%%%%%%

Our main results are summarized in Fig.~\ref{fig:PhaseDiagram}(c). The  magnetic phase 
diagram at low temperatures is divided into three regions. At low fields, and very low 
temperatures, the spins show a quasi-long-range order (QLRO) in the $XY$ plane in 
a canted antiferromgnetic (single-$q$) structure which breaks the translational and orientational
symmetries of the lattice. By increasing the temperature while keeping fixed
the value of the magnetic field, we expect a Kosterlitz-Thouless (KT) transition to
a phase in which the internal structure melts down. This intermediate phase has 
the $Z_3$ spatial symmetry broken, similar to what was found in \cite{Mulder2010}.

%%%%%%%%%%%%%%%%%%%%%%%%%%%%%%%%%%%%%%%%%%%%%%%%%%%%%%%%%%%%%%%%%%%%%%%%%%%%%%%%%%%%%%%%%%%%%%%%%%%
%%%%%%%%%%%%%%%%%%%%%%%%%%%%%%%%%%%%%%%%%%%%%%%%%%%%%%%%%%%%%%%%%%%%%%%%%%%%%%%%%%%%%%%%%%%%%%%%%%%

\begin{figure}[t!]
 \begin{center}
 \includegraphics[height=5cm]{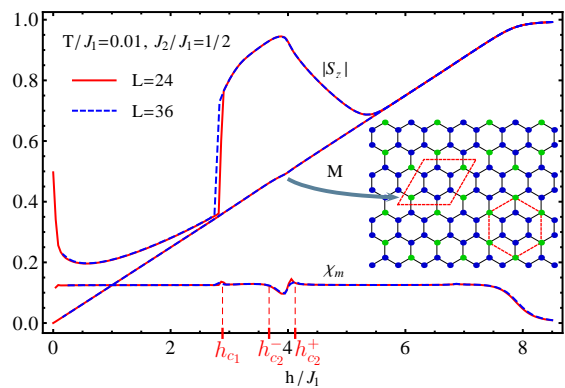}
 \end{center}
\caption{(color online) Magnetization curve, absolute value of total $S_z$ and  susceptibility as a function of
the magnetic field for $J_2/J_1=1/2$ for two system sizes ($N=2*L^2$).
Inset:  $S_z$ spin configuration, $uuuuuudd$ state (collinear $3-1$), on the $M/M_{sat}=1/2$ quasi-plateau at $T/J_1=0.01$.
 The blue and green dots correspond to a spin fully polarized along the magnetic field and in the opposite direction
 respectively.}
\label{fig:M-Chi-Szvsh}
\end{figure}

%%%%%%%%%%%%%%%%%%%%%%%%%%%%%%%%%%%%%%%%%%%%%%%%%%%%%%%%%%%%%%%%%%%%%%%%%%%%%%%%%%%%%%%
When the value of the magnetic field exceeds $h_{c_1}/J_1\simeq 2.9$, we find
a first order transition into an intermediate phase similar to the 2--1--1 phase observed
in the pyrochlore  antiferromagnet ZnCr$_2$O$_4$ \cite{211experimental} which transforms at 
$h_{c_2}/J_1 =4$ into the collinear 3--1 structure via a second-order transition. Although 
in all high field phases the unit cell of the system have 8 spins, we shall use the same 
terminology as in \cite{211experimental} to emphasize the magnetic structure for each frustrated 
four-spin block. The collinear 3--1 state corresponds to the magnetization plateau with 
$M/M_{\rm sat}=1/2$. It preserves the $XY$ rotational symmetry about the field direction but breaks
in a special way the discrete $Z_4$ lattice translational symmetry (see the inset in Fig.~\ref{fig:M-Chi-Szvsh}).
For this value of the magnetization (magnetic field)  increasing  temperature
produces a phase transition  into the paramagnetic phase via
a continuous transition in the universality class of the $Z_4$ clock model.
It is interesting to notice that the 2--1--1 state has a supplementary $Z_3$ broken symmetry with respect to the 3--1 state. One would then expect again a 3-states Potts transition by keeping the temperature fixed and increasing the magnetic field in order to pass from the 2--1--1 to the 3--1 state.
%%%%%%%%%%%%%%%%%%%%%%%%%%%%%%%%%%%%%%%%%%%%%%%%%%%%%%%%%%%%%%%%%%%%%%%%%%%%%%%%%%%%%%%%%%%%%%%%%%

By increasing the applied magnetic field beyond the 1/2 plateau, we enter into the high-field phase
where a 3--1 state known previously for the pyrochlore antiferromagnet \cite{211experimental} becomes
stable.  The classical configurations for the spins are now canted again, we have broken translational
symmetry due to the long range order in the $z$ component of the spins and we  recover quasi-long-range order
for the $XY$ components. Note that, in contrast to the  low field phase, the lattice orientational symmetry is unbroken.
%

%%%%%%%%%%%%%%%%%%%%%%%%%%%%%%
To see all this in detail, let us now introduce the model. The spin Hamiltonian is given by
\ba
H=J_{1}\!\sum_{<i,j>_{1}}\op{S}_{i}\cdot\op{S}_{j}+J_{2}\sum_{<i,j>_{2}}\op{S}_{i}\cdot\op{S}_{j}-h\sum_{i}S^{z}_{i}
\label{eq:Hamil}
\ea
where $<i,j>_{1}$ and $<i,j>_{2}$ denote nearest and next nearest neighbors respectively, 
$J_1,J_2>0$ and $h$ is the external magnetic field.

Let us briefly discuss the magnetically ordered phases of the model  (\ref{eq:Hamil}). For small 
diagonal exchange $J_2/J_1<1/6$, classical spins form the Neel state. For $1/6<J_2/J_1\leq 1/2$ 
the classical ground states are degenerate spirals forming a closed contour around $\Gamma$-point, while for $1/2<J_2/J_1$, the closed contours are centered around K and K'-points \cite{Mulder2010}.

%%%%%%%%%%%%%%%%%%%%%%%%%%%%%%%%%%%%%%%%%%%%%%%%%%%%%%%%%%%%%%%%%%%%%%%%%%%%%%%%%%%%%%%%%%%%%%%%%%%

We now focus on the highly frustrated point $J_2/J_1=1/2$, where the Hamiltonian  can be written, 
up to a constant term, as a sum over elementary tetrahedrons that we label as $\vee$ and $\wedge$ 
respectively (see Fig.~\ref{fig:PhaseDiagram}b):
\ba
\nn
H=\frac{J_1}{4}\sum_{\Omega=\vee,\wedge}
\left(\op{S}^2_{\Omega}-\frac{1}{J_1}{\bf h}\cdot\op{S}_{\Omega}\right),
\label{eq:Hplaquette}
\ea
where $\op{S}_{\vee}=\sum_{i\in \vee} \op{S}_i$, $\op{S}_{\wedge}=\sum_{i\in \wedge} \op{S}_i$
and ${\bf h}=h\,\hat{z}$.
By minimizing the energy on each $\wedge$ and $\vee$, one obtains the constraint:
$\op{S}_{\vee}=\op{S}_{\wedge}= {\bf h}/(2J_1)$.
The classical ground state is obtained when this constraint is satisfied in every block and  presents only the typical global rotation as a degeneracy. The saturation field $h_{s}$ is determined by the condition $S^z_{\wedge} = 4$ and  $S^z_{\vee} = 4$ which gives $h_{s}=8\,J_1$. At this value all the spins are aligned with the $z-$axis.

Monte Carlo simulations have been performed using the standard Metropolis algorithm in
combination with the microcanonical over-relaxation steps, see \cite{Zhitomirsky08} for
further details. Periodic boundary conditions were implemented for $N=2\times L^2$  site clusters
with $L=24$--$72$.  At every magnetic  field or temperature we discarded $10^5$  Monte Carlo steps
(MCS) for initial relaxation and data were collected during subsequent $2\cdot 10^5$ MCS. The error  bars were estimated
from  20 independent runs initialized by different random numbers.

Let us now discuss various physical quantities used to clarify different phases
and corresponding transitions. In the first place, we calculate the magnetization, susceptibility and absolute value of $S_z$ defined as
\ba
M=\frac{1}{N}\sum_{i=1}^{N}S^{z}_i,   &\
\chi_m={\displaystyle \frac{dM}{dh}}, &\
|S_z|=\frac{1}{N}\sum_{i=1}^{N}|S^{z}_i|.
\ea
In Fig.~\ref{fig:M-Chi-Szvsh} we show the magnetization curve, susceptibility and absolute value of $S_z$  as a function of the external field at temperature $T=0.01\,J_1$.

The susceptibility $\chi_m$ shows a dip around $h=h_{c_2}= 4\,J_1$ which indicates the presence of a quasi-plateau phase. In the same region the absolute value of $|S_z|$, which measures how ``collinear'' is the magnetic configuration, is close to one and therefore  the magnetic phase established  is a ``collinear phase''.
The situation is completely different for small fields, $|S_z|$ is smaller that $1/2$ and then the phase corresponds to a canted 2-2 antiferromagnetic (AF). Both regions are separated by a big jump in $|S_z|$ around $h=h_{c_1}\simeq2.9\,J_1$ indicating a first order phase transition.

The previous results suggest that the low field phase is continuously connected with the zero field case studied in \cite{Mulder2010}. Fluctuations select a commensurate wave vector corresponding to the $M$-point in the Brillouin zone (BZ). It has residual triple degeneracy. At zero magnetic field the selected structure is described by a single wavevector.  To detect this single-$q$-paramagnetic phase transition we introduce a local complex order parameter $\Delta^{\alpha\beta}(\vr)$ \cite{Mulder2010} and its averages as
\ba
\Delta^{\alpha\beta}(\vr)\!\!\! &=&\!\!\!\frac{1}{2}S^{\alpha}_A(\vr)\left[S^{\beta}_B(\vr)\! +\!
\omega \,S^{\beta}_B(\vr\!+\!{\bf b})\!+\!\omega^2  \, S^{\beta}_B(\vr\!-\!{\bf a}\!+\!{\bf b})\right]\nn\\
\label{eq:OP}
\Delta&=&\Big|\frac{1}{N_c}\sum_{\vr\in A}\Delta^{xx}(\vr)+\Delta^{yy}(\vr)+\Delta^{zz}(\vr)\Big|\\
\label{eq:OPperp}
\Delta_\perp&=&\Big|\frac{1}{N_c}\sum_{\vr\in A}\Delta^{xx}(\vr)+\Delta^{yy}(\vr)\Big|\\
\label{eq:OPz}
\Delta_{zz}&=&\Big|\frac{1}{N_c}\sum_{\vr\in A}\Delta^{zz}(\vr)\Big|
\ea
where the sum over $\vr$ runs over one of the two sublattices (say $A$),
$\omega=e^{2\pi i/3}$, $\alpha=x,y,z$ and ${\bf a}$, ${\bf b}$ are the primitive translation vectors of
 the direct lattice (See Fig. \ref{fig:PhaseDiagram}). In the previous definition we have normalized to 1 the case of a perfect ''two-up, two-down'' collinear configuration (only achievable at $T=0$).\\

\begin{figure}[t!]
\begin{center}
\includegraphics[height=4.4cm]{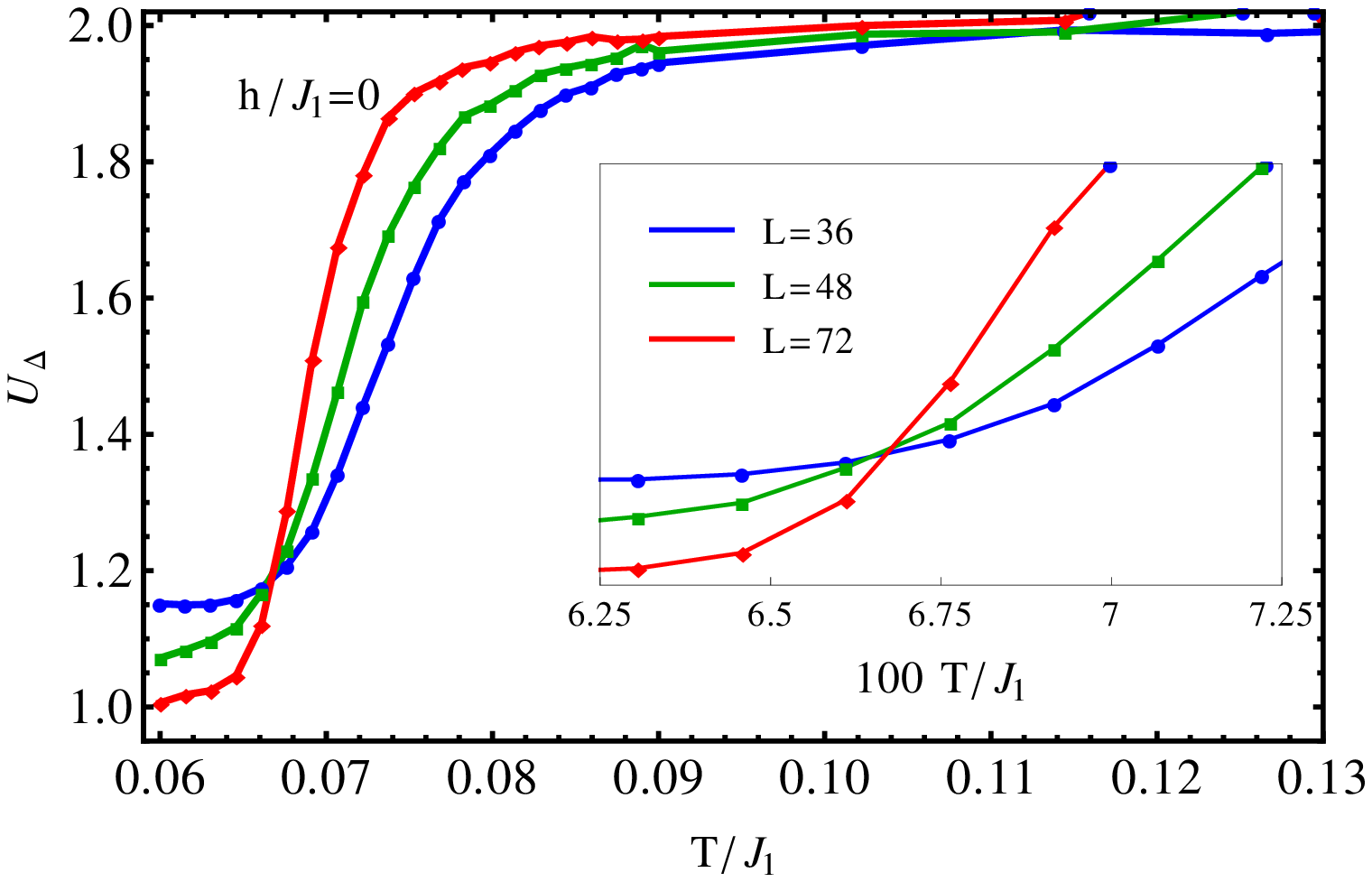}
\includegraphics[height=4.4cm]{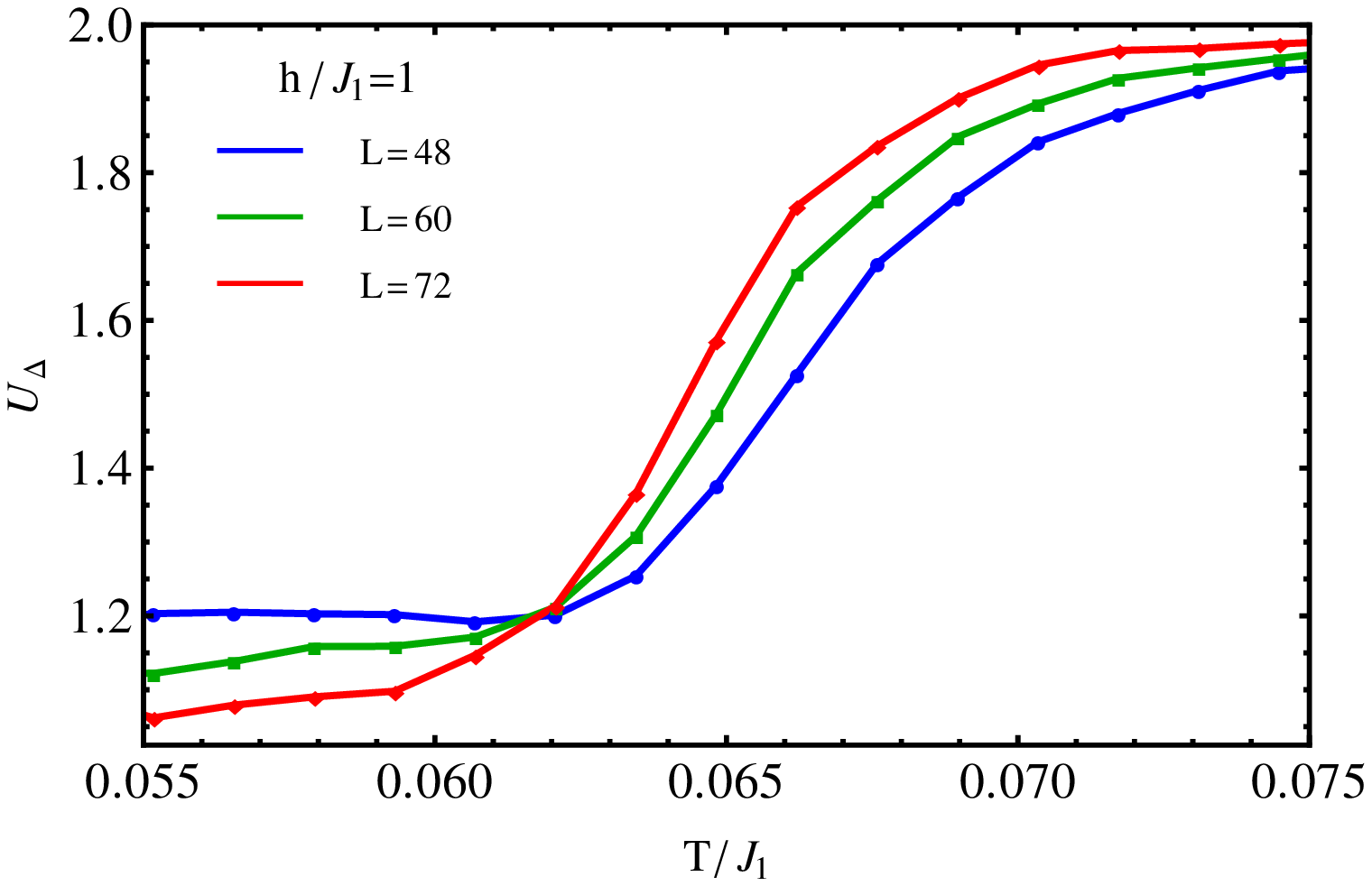}
\end{center}
\caption{Binder cumulant associated with the order parameter $\Delta$ as a function of temperature $T$ showing the transition between the paramagnetic phase
and the single-$q$ phase  at $h/J_1 = 0$ (Top) and $h/J_1=1$ (bottom), for system sizes up to $L=72$.} \label{fig:chi-U-vs-T}
\end{figure}

Following the standard procedure, the second-order transition between a paramagnetic phase (large-$T$) and a single-$q$ phase  may be located by the crossing point of the corresponding  Binder cumulant $U_{\Delta}$ measured for different clusters. We have used instantaneous values of (\ref{eq:OP}) %, (\ref{eq:OPperp}) and (\ref{eq:OPz})
to measure the  susceptibility $\chi_{\Delta}$ and the Binder cumulant $U_{\Delta}$ associated with this order parameter defined as
\ba \chi_{\Delta}=\frac{N_c}{T}\langle
(\Delta)^2\rangle,&\quad&U_{\Delta}=\frac{\langle(\Delta)^4\rangle}{\langle
(\Delta)^2\rangle}
\label{eq:chi-and-U} \ea
where $N_c$ is the number of unit cells. We illustrate this method in the top panel of Fig.~\ref{fig:chi-U-vs-T} for the transition between the paramagnetic state and the single-$q$ state.
The alternative approach is to study the susceptibility since the critical exponent  $\eta$ is known precisely, $\eta=4/15$ \cite{CK}. In the
critical region the susceptibility scales as
\ba \chi_{\Delta}&=&L^{2-\eta}f(|\tau|L^{1/\nu}),\,\tau=1-T/T_c. \ea
Hence, the normalized susceptibility $\chi_\Delta/L^{2-\eta}$ becomes size-independent at $\tau=0$ and curves for different $L$ plotted as functions of $T$ exhibit a crossing point, similar to the behavior observed for the Binder cumulant.

%
%%%%%%%%%%%%%%%%%%%%%%%%%%%%%%%%%%%%%%%%%%%%%%%%%%%%%%%%%%%%%%%%%%%%%%%%%%%%%%%%%%%%%%%%%%%%%%%%%%%
%

In the case of non zero field we find three different regions, as it is schematically depicted in Fig.
(\ref{fig:PhaseDiagram}):\\[0.5mm]
(i)\ For $h<h_{c_1}$ we have a similar situation as for zero field, namely at fixed magnetic field and coming from high temperature, a paramagnetic-$Z_{3}$ transition occurs. Decreasing further the temperature one then should encounter a KT transition to a canted 2-2 AF single-$q$ QLRO in the $XY$ spin components. \\[0.5mm]
(ii)\ For $h_{c_1}<h<h_{c_2}$  we have a finite-temperature transition related with the breaking 
of $Z_4 \times Z_3$ spatial symmetry. Again, at low temperatures one should find a KT transition to a QLRO phase in the $XY$ spin plane corresponding to a coplanar configuration similar to that found in \cite{211experimental} for the pyrochlore lattice, which we dubbed the 2--1--1 phase.
Note that close to $h_{c_1}$, thermal fluctuations select a collinear configuration for the spins which dramatically decrease the effective spin stiffness in the $XY$ plane when entering in the phase II. Then one expects
the KT transition temperature in region II to be lower than the one of region I, as schematically depicted in Fig. (\ref{fig:PhaseDiagram}). \\[0.5mm]
(iii)\ For $h>h_{c_2}$, coming from the high temperature phase, we encounter a continuous transition to a phase with broken $Z_{4}$ translational symmetry, see  the inset in Fig.~\ref{fig:M-Chi-Szvsh}. 
This 3--1 phase as well as the 2--1--1 phase are characterized by mixing of three wave-vectors
(triple-$q$ structures).
Again, at even lower temperatures, we expect a KT transition to a QLRO state where the spins adopt a planar configuration in which
three spins are pointing in the same direction and the $XY$ component of the remaining spin
compensates the sum of the three first ones.

\begin{figure}[t!]
\begin{center}
\includegraphics[height=5.5cm]{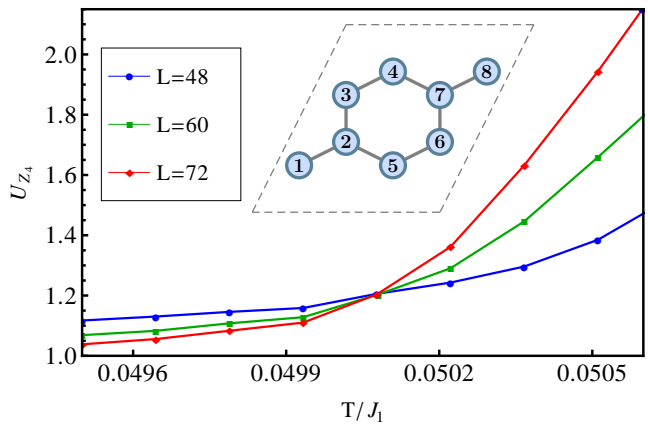}
\end{center}
\caption{(color online) Binder Cumulant corresponding to the $Z_4$ order parameter at $h/J_1=4.2$ defined in the main text.
The transition between the paramagnetic phase and the $3-1$ state is clearly observed.}
\label{fig:UvsT(H3.9)}
\end{figure}

Thermal fluctuations have a strong effect in fields around $\frac{1}{2}h_s$, where they stabilize 
a collinear 3--1 state and there is a symmetry breaking related to this selection, as we explain now.
At this point the spin pattern consists of 8 spins per unit cell (Fig.~\ref{fig:M-Chi-Szvsh}).
We rewrite the coordinates of the 8 spins as we show in the inset of Fig. \ref{fig:UvsT(H3.9)}
and introduce  the  following $Z_4$ order parameter
\ba
m^z&=&\frac{1}{N}\sum_{\sumauno}\sum_{n=1}^{4}e^{i\frac{\pi}{2}(n-1)}S^z_{\sumauno,n}+e^{i\frac{\pi}{2}(4-n)}S^z_{\sumauno,n+4}.
\ea
Using this order parameter one can construct the corresponding Binder cumulant $U_{Z_4}$ in the usual way.
%
%\ba
%U_{Z_4}&=&\frac{\langle (m^z)^4\rangle}{\langle (m^z)^2\rangle^2},
%\ea
%
The results are shown in Fig.~\ref{fig:UvsT(H3.9)} (at  $h/J_1=4.2$), measured for different
cluster sizes at the transition between the paramagnetic and the $Z_4$ symmetry
breaking collinear $3-1$ state.
Results for various temperatures and magnetic field scans are summarized
in the phase diagram  presented in Fig.~\ref{fig:PhaseDiagram}(c)

To summarize, we have studied the phase diagram of a strongly frustrated classical $J_1$--$J_2$ Heisenberg
antiferromagnet on a honeycomb lattice in a magnetic field. We have found a very rich low temperature
phase diagram showing three non-trivial regions characterized by different broken lattice symmetries, as
summarized in Fig. (1).

In order of increasing magnetic field at fixed (low) temperature, one first encounters a first order transition
triggered from the low field phase (I), which breaks the orientational $Z_3$ symmetry,
into the intermediate field phase (II) where a supplementary $Z_4$ symmetry, related to lattice translations, is broken.
By increasing the field further, a continuous transition  to phase (III) occurs, which provides an
example of (lattice) symmetry restoration.

All these discrete broken symmetry phases should coexist with the corresponding QLRO phases, after a KT
transition occurs at lower temperatures.

We have explicitly numerically checked that the transition from phase (III) to the paramagnetic phase is
continuous and one should then in principle expect continuously varying exponents associated with the universality class
of the $Z_4$ Potts model \cite{Z4}.
The richness of this phase diagram illustrates the importance of frustrated and competing interactions and the
onset of fluctuation mechanisms in the selection of the low energy configurations.

The present study may be relevant in the study of different compounds that are described by the frustrated hexagonal
Heisenberg model, such as $\rm Bi_3Mn_4O_{12}(NO_3)$ \cite{Matsuda10} already mentioned and the
family of compounds $\rm BaM_2(XO_4)_2$ with M = Co, Ni and X = P, As, which consist of
magnetic ions M arranged in weakly coupled frustrated honeycomb lattices with spin S = 1/2 for
Co and S = 1 for Ni \cite{materials}. In the case of materials with spins higher than $1/2$, magnetic field
experiments could unravel some of the structures found in the present paper. Last but not least, the controlled setup of optical lattices
for cold atoms would allow to create arbitrary lattice structures as well as to tune the interactions \cite{coldatoms}.

{\it Acknowledgements.}---%
CAL and PP acknowledge very fruitful discussion with N. Laflorencie.
HDR, DCC and CAL  are partially supported by CONICET (PIP 1691) and ANPCyT
(PICT 1426). PP  and MEZ acknowledge support by the Agence Nationale de la Recherche
under grants No.\ ANR 2010 BLANC 0406-0 and  No.\ ANR-09-Blanc-0211, respectively.

\end{document}